\documentclass[pra,twocolumn,superscriptaddress,amsfonts,amssymb,amsmath,nofootinbib]{revtex4-2}

\newcommand{\exporttikz}{0}
\newcommand{\dowordcount}{0}

\usepackage{mathtools,mymath}
\usepackage{braket}
\usepackage{ifthen}
\usepackage{xcolor}
\usepackage{booktabs}
\usepackage{array}
\newcolumntype{C}[1]{>{\centering\let\newline\\\arraybackslash\hspace{0pt}}m{#1}}
\newcolumntype{L}[1]{>{\let\newline\\\arraybackslash\hspace{0pt}}m{#1}}
\usepackage[normalem]{ulem}

\newtheorem{theorem}{Theorem}
\newtheorem{definition}{Definition}
\newtheorem{lemma}{Lemma}

\usepackage[colorlinks,urlcolor=blue,linkcolor=black,citecolor=blue]{hyperref}
\usepackage[capitalize,nameinlink]{cleveref}
\usepackage[caption=false]{subfig}

\ifnum\exporttikz>0
  \usepackage{tikz}
  \usepackage{pgfplots}
  \usepackage{myqcircuit}
  \usetikzlibrary{calc}
  \usetikzlibrary{positioning}
  \usepgfplotslibrary{statistics}
  \pgfplotsset{
    compat=newest,
    table/header=false,
    tick label style={font=\footnotesize},
    label style={font=\footnotesize},
    legend style={font=\scriptsize},
    legend cell align=left,
    legend style={draw=none},
  }
  \usepgfplotslibrary{external}
\else
  \usepackage{tikzexternal}
  \tikzsetexternalprefix{fig/}
\fi
\newcommand{\figname}[1]{\tikzsetnextfilename{#1}}

\newcommand{\datfile}[1]{fig/dat/#1.dat}

\ifnum\exporttikz>0
  \makeatletter
    \pgfplotsset{
      boxplot/hide outliers/.code={\def\pgfplotsplothandlerboxplot@outlier{}}
    }
  \makeatother
\fi

\usepackage{environ}
\ifnum\dowordcount=1
  \NewEnviron{myequation}{}
  \NewEnviron{myequation*}{}
  \NewEnviron{myalign}{}
  \NewEnviron{myalign*}{}
\else
  \NewEnviron{myequation}{\begin{equation}\BODY\end{equation}}
  \NewEnviron{myequation*}{\begin{equation*}\BODY\end{equation*}}
  \NewEnviron{myalign}{\begin{align}\BODY\end{align}}
  \NewEnviron{myalign*}{\begin{align*}\BODY\end{align*}}
\fi

\DeclareMathOperator{\swap}{SWAP}
\DeclareMathOperator{\polylog}{polylog}
\DeclareMathOperator{\poly}{poly}
\DeclareMathOperator*{\myprod}{\Pi}
\DeclareMathOperator*{\mysum}{\Sigma}

\definecolor{mycolor1}{rgb}{0.00000,0.44700,0.74100}%
\definecolor{mycolor2}{rgb}{0.85000,0.32500,0.09800}%
\definecolor{mycolor3}{rgb}{0.92900,0.69400,0.12500}%
\definecolor{mycolor4}{rgb}{0.49400,0.18400,0.55600}%
\newcommand{\myTDFIMboxplot}[7]{%
\ifthenelse{#1=1}
{
\def\myylabel{relative error}
}
{
\def\myylabel{repetitions until success}
}
\begin{tikzpicture}
\begin{axis}[
  width=\columnwidth,%
  height=0.75\columnwidth,%
  boxplot/draw direction=y,%
  boxplot/box extend=0.2,%
  boxplot/hide outliers,
  xlabel=$s$ spins,%
  ylabel=\myylabel,%
  xmin=1.25,xmax=10.75,%
  xtick={2,3,4,5,6,7,8,9,10},%
  ymin=0,ymax=0.05,%
  ytick={0,0.01,...,0.05},%
  yticklabel style={
        /pgf/number format/fixed,
        /pgf/number format/precision=4
  },%
  scaled y ticks=false,%
  legend style={at={(0.5,0.99)},anchor=north,legend columns=-1},%
  legend entries={$\sigma_{x,z}$,LCU,$\sigma_{x,z}$ and LCU},%
]
\addlegendimage{thick,mycolor1,mark=square,mark size=1.5pt}
\addlegendimage{thick,mycolor2,mark=o,mark size=1.5pt}
\addlegendimage{thick,mycolor3,mark=diamond}
\def\XosA{-0.25}
\def\XosB{0}
\def\XosC{0.25}
\addplot[thick,boxplot,boxplot/draw position=2+\XosA,mycolor1]%
  table[y index=0, col sep=tab] {\datfile{#2}};
\addplot[thick,boxplot,boxplot/draw position=2+\XosB,mycolor2]%
  table[y index=0, col sep=tab] {\datfile{#3}};
\addplot[thick,boxplot,boxplot/draw position=2+\XosC,mycolor3]%
  table[y index=0, col sep=tab] {\datfile{#4}};

\ifthenelse{#1=1}{
\addplot[thick,densely dotted, mark=square, mark size=1.5pt, mark options={solid,fill=mycolor1}, color=mycolor1]%
  table[x expr=\thisrowno{0}+\XosA, y index=1, col sep=tab] {\datfile{#5}};
  \addplot[thick,densely dotted, mark=o, mark size=1.5pt, mark options={solid,fill=mycolor2}, color=mycolor2]%
  table[x expr=\thisrowno{0}+\XosB, y index=1, col sep=tab] {\datfile{#6}};
  \addplot[thick,densely dotted, mark=diamond, mark options={solid,fill=mycolor3}, color=mycolor3]%
  table[x expr=\thisrowno{0}+\XosC, y index=1, col sep=tab] {\datfile{#7}};
}{
\ifthenelse{#1=2}{
\addplot[only marks,mark=diamond*, mark options={fill=black}, color=black]%
  table[x index=0, y index=1, col sep=tab] {\datfile{#5}};
}{}}

\addplot[thick,boxplot,boxplot/draw position=3+\XosA,mycolor1]%
  table[y index=1, col sep=tab] {\datfile{#2}};
\addplot[thick,boxplot,boxplot/draw position=3+\XosB,mycolor2]%
  table[y index=1, col sep=tab] {\datfile{#3}};
\addplot[thick,boxplot,boxplot/draw position=3+\XosC,mycolor3]%
  table[y index=1, col sep=tab] {\datfile{#4}};

\addplot[thick,boxplot,boxplot/draw position=4+\XosA,mycolor1]%
  table[y index=2, col sep=tab] {\datfile{#2}};
\addplot[thick,boxplot,boxplot/draw position=4+\XosB,mycolor2]%
  table[y index=2, col sep=tab] {\datfile{#3}};
\addplot[thick,boxplot,boxplot/draw position=4+\XosC,mycolor3]%
  table[y index=2, col sep=tab] {\datfile{#4}};

\addplot[thick,boxplot,boxplot/draw position=5+\XosA,mycolor1]%
  table[y index=3, col sep=tab] {\datfile{#2}};
\addplot[thick,boxplot,boxplot/draw position=5+\XosB,mycolor2]%
  table[y index=3, col sep=tab] {\datfile{#3}};
\addplot[thick,boxplot,boxplot/draw position=5+\XosC,mycolor3]%
  table[y index=3, col sep=tab] {\datfile{#4}};

\addplot[thick,boxplot,boxplot/draw position=6+\XosA,mycolor1]%
  table[y index=4, col sep=tab] {\datfile{#2}};
\addplot[thick,boxplot,boxplot/draw position=6+\XosB,mycolor2]%
  table[y index=4, col sep=tab] {\datfile{#3}};
\addplot[thick,boxplot,boxplot/draw position=6+\XosC,mycolor3]%
  table[y index=4, col sep=tab] {\datfile{#4}};

\addplot[thick,boxplot,boxplot/draw position=7+\XosA,mycolor1]%
  table[y index=5, col sep=tab] {\datfile{#2}};
\addplot[thick,boxplot,boxplot/draw position=7+\XosB,mycolor2]%
  table[y index=5, col sep=tab] {\datfile{#3}};
\addplot[thick,boxplot,boxplot/draw position=7+\XosC,mycolor3]%
  table[y index=5, col sep=tab] {\datfile{#4}};

\addplot[thick,boxplot,boxplot/draw position=8+\XosA,mycolor1]%
  table[y index=6, col sep=tab] {\datfile{#2}};
\addplot[thick,boxplot,boxplot/draw position=8+\XosB,mycolor2]%
  table[y index=6, col sep=tab] {\datfile{#3}};
\addplot[thick,boxplot,boxplot/draw position=8+\XosC,mycolor3]%
  table[y index=6, col sep=tab] {\datfile{#4}};

\addplot[thick,boxplot,boxplot/draw position=9+\XosA,mycolor1]%
  table[y index=7, col sep=tab] {\datfile{#2}};
\addplot[thick,boxplot,boxplot/draw position=9+\XosB,mycolor2]%
  table[y index=7, col sep=tab] {\datfile{#3}};
\addplot[thick,boxplot,boxplot/draw position=9+\XosC,mycolor3]%
  table[y index=7, col sep=tab] {\datfile{#4}};

\addplot[thick,boxplot,boxplot/draw position=10+\XosA,mycolor1]%
  table[y index=8, col sep=tab] {\datfile{#2}};
\addplot[thick,boxplot,boxplot/draw position=10+\XosB,mycolor2]%
  table[y index=8, col sep=tab] {\datfile{#3}};
\addplot[thick,boxplot,boxplot/draw position=10+\XosC,mycolor3]%
  table[y index=8, col sep=tab] {\datfile{#4}};

\end{axis}
\end{tikzpicture}
}

\begin{document}

\ifnum\dowordcount=0

\title{Approximate Quantum Circuit Synthesis using Block-Encodings}

\author{Daan Camps}
\email{dcamps@lbl.gov}
\affiliation{Computational Research Division, Lawrence Berkeley National Laboratory, Berkeley, CA 94720, USA}
\author{Roel Van Beeumen}
\email{rvanbeeumen@lbl.gov}
\affiliation{Computational Research Division, Lawrence Berkeley National Laboratory, Berkeley, CA 94720, USA}

\date{\today}

\begin{abstract}
One of the challenges in quantum computing is the synthesis of unitary
operators into quantum circuits with polylogarithmic gate complexity.
Exact synthesis of generic unitaries requires an exponential
number of gates in general.
We propose a novel approximate quantum circuit synthesis technique by relaxing the
unitary constraints and interchanging them for ancilla qubits via block-encodings.
This approach combines smaller block-encodings, which are
easier to synthesize, into quantum circuits for larger operators.
Due to the use of block-encodings, our technique is not limited to unitary
operators and can also be applied for the synthesis of arbitrary operators.
We show that operators which can be approximated by a canonical polyadic
expression with a polylogarithmic number of terms can be synthesized with polylogarithmic gate complexity
with respect to the matrix dimension.
\end{abstract}

\maketitle

\fi

\section{Introduction}
Quantum computing holds the promise of speeding up computations
in a wide variety of fields \cite{Nielsen:2011:QCQI}.
After early breakthroughs such as Shor's algorithm \cite{shor1994} for factoring
and Grover's algorithm \cite{grov1996} for searching,
there have been substantial developments in various quantum algorithms over the past two decades.
Noteworthy are the quantum walk algorithm of Szegedy \cite{quant-ph/0401053,szeg2004},
 and the quantum linear systems
algorithm by Harrow, Hassidim, and Loyd \cite{haha2009}.
These developments have lead to quantum linear systems \cite{chko2017} and Hamiltonian simulation \cite{bech2015} algorithms inspired by quantum walks.
A unifying framework called the quantum singular value transformation, which combines
the notion of qubitization \cite{loch2019} and quantum signal processing \cite{loch2017} by Low and Chuang,
was recently proposed by Gily\'en et al.~\cite{gisu2018,gisu2019}.
The quantum singular value transformation can describe all aforementioned quantum algorithms except factoring.
Besides that, it has sparked an interest in the use of block-encodings since they can directly be used as input for a quantum singular value transformation. A block-encoding is the embedding of a --not necessarily unitary-- operator as the leading principal block in a larger unitary
\begin{myequation}
U = \begin{bmatrix} A/\alpha & *\, \\ * & *\, \end{bmatrix}
\ \ \Longleftrightarrow \ \
A = \alpha \left(\bra0 \otimes \eye\right) U \left(\ket0 \otimes \eye\right),
\end{myequation}
where $*$ indicate arbitrary matrix elements.

In this paper, we propose the use of block-encodings, not as a building block for quantum algorithms, but as a technique for \emph{approximate} quantum circuit synthesis and, more generally, the synthesis of arbitrary operators into quantum circuits.
One of the major challenges on noisy intermediate-scale quantum (NISQ)
devices is the limited circuit depth \cite{pres2018}.
In general, exact synthesis of generic unitary operators requires
exponentially many quantum gates \cite{Kitaev:2002:CQC,shbu2006,dani2006}.
The noise in NISQ devices limits the circuit depth but also relaxes the
need for exact synthesis.
In other words, we only need to approximate the action of some $n$-qubit
operator up to an error proportional to the noise level.
A polynomial dependence of the circuit depth on $n$ is necessary to obtain
efficient quantum circuits.
Examples of other approximate synthesis approaches have been proposed in
\cite{pasv2014,boro2015,mamo2016,khlr2019,yose2020}.

We show that, under certain assumptions, an efficient quantum circuit can be devised if the operator can be $\epsilon$-approximated by a canonical polyadic (CP) expression \cite{koba2009,hitc1927} with a number of terms that depends polylogarithmically on the operator dimension.
We denote these by \emph{PLTCP matrices}.
CP decompositions have found applications in many scientific disciplines because they can often be computed approximately using optimization algorithms. However, their calculation is an NP-hard problem in general.
We also demonstrate that the class of operators that we can efficiently synthesize is a linear combination of terms with Kronecker product structure, which is more general than standard CP decompositions.
We call these \emph{CP-like} decompositions.

The proposed technique uses two operations to efficiently combine block-encodings: the Kronecker product of block-encodings and a linear combination of block-encodings.
This allows us to combine block-encodings of small matrices into quantum circuits for larger operators.
We show that in practice the scheme requires at most a logarithmic number of ancilla qubits, 
study the relation between the errors on the individual encodings and the overall circuit,
and analyze the CNOT complexity of the circuits.
Finally, we show three examples of 
non-unitary operators that naturally have a CP-like structure and can efficiently be encoded using the proposed technique.

\section{Block-encodings}

Since an $n$-qubit quantum circuit performs a unitary operation, non-unitary
operations cannot directly be handled by quantum computers.
One way to overcome this limitation is by encoding the non-unitary matrix into
a larger unitary one, so called \emph{block-encoding}~\cite{gisu2018,gisu2019}.
We define an \emph{approximate} block-encoding of an operator on $s$ signal qubits, $\eA_s$, in a unitary $\eU_n$ on $n$ qubits as follows.

\begin{definition}
\label{def:BE}
Let $a,s,n \inN$ such that $n = a + s$, and $\epsilon\inR^+$.
Then an $n$-qubit unitary $\eU_n$ is an $(\alpha,a,\epsilon)$-block-encoding of an $s$-qubit operator $\eA_s$ if
\begin{myequation}
\tilde \eA_s = \left(\bra{0}^{\otimes a} \otimes \eI_{s} \right) \eU_n
               \left(\ket{0}^{\otimes a} \otimes \eI_{s} \right),
\end{myequation}
and
\(
\normtwo[\big]{\eA_s - \alpha \tilde \eA_s} \leq \epsilon.
\)
\end{definition}

The parameters $(\alpha, a, \epsilon)$ of the block-encoding are, respectively, the \emph{subnormalization factor} to encode matrices of arbitrary norm, the number of \emph{ancilla} qubits, and the \emph{error} of the block-encoding.
Since $\normtwo{\eU_n} = 1$, we have that $\normtwo{\tilde \eA_s} \leq 1$ and $\normtwo{\eA_s} \leq \alpha + \epsilon$.
Note that every unitary $U_s$ is already a $(1,0,0)$-block-encoding of itself and every non-unitary matrix $\eA_s$ can be embedded in a $(\normtwo{\eA_s},1,0)$-block-encoding \cite{QI:Alber:2001}.
This does not guarantee the existence of an efficient quantum circuit.

\begin{figure}[hbtp]
\centering\ifnum\dowordcount=0
\figname{circuit-Un}%
\begin{tikzpicture}[on grid]
\pgfkeys{/myqcircuit, layer width=10mm, row sep=3mm, source node=qwsource}
\newcommand{\qwstart}{1}
\newcommand{\qwend}{1}
\qwire[start node=qwsource, end node=qw1e, style=thin,            index=1, start layer=\qwstart, end layer=\qwend, label=$\ket{0}^{\otimes a}$]
\qwire[start node=qw3s,       end node=qw3e, style=markedwire, index=3, start layer=\qwstart, end layer=\qwend, label=$\ket{\psi_s}$ ]
\multiqubit[label=$\eU_n$, layer=1.1, start index=1, stop index=3, node=Un]
\singlequbit[style=bundle, index=1, layer=0.5]
\singlequbit[style=bundle, index=3, layer=0.5]
\measurement[layer=2, index=1, result=$0$]
\node[node distance=0.5*\layerwidth] () [right of=qw3e] {$\tilde \eA_s \ket{\psi_s}$};
\end{tikzpicture}
\fi
\caption{Quantum circuit for $\eU_n$. The thick quantum wire carries the \emph{signal} qubits, the other are the \emph{ancilla} qubits. If the ancilla register is measured in the zero state, the signal register is in the desired state $\tilde \eA_s \ket{\psi_s}$.\label{fig:BE}}
\end{figure}
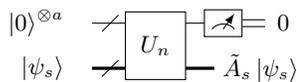

An equivalent interpretation of \Cref{def:BE} is that $\tilde \eA_s$ is the partial trace of $\eU_n$ over the zero state of the ancilla space. This naturally partitions the Hilbert space $\cH_n$ into $\cH_a \otimes \cH_s$.
Given an $s$ qubit signal state, $\ket{\psi_s} \in \cH_s$,
the action of $\eU_n$ on $\ket{\psi_n} = \ket{0}^{\otimes a} \otimes \ket{\psi_s} \in \cH_n$ becomes
\begin{myequation}
\eU_n \ket{\psi_n} = \ket{0}^{\otimes a} \otimes \tilde \eA_s \ket{\psi_s}
+ \sqrt{1 - \normtwo{ \tilde \eA_s \ket{\psi_s}}^2} \ket{\phi^{\perp}_n},
\end{myequation}
with
\begin{myequation}
\left(\bra{0}^{\otimes a} \otimes \eI_{s}\right) \ket{\phi^{\perp}_n} =0,
\qquad \normtwo[\big]{\ket{\phi^{\perp}_n}} =1,
\end{myequation}
and $\ket{\phi^{\perp}_n}$ the normalized state for which the ancilla register has a state orthogonal to $\ket{0}^{\otimes a}$.
By construction, we see that a partial measurement of the ancilla register projects out $\ket{\phi^{\perp}_n}$ and results in
$(\ket{0}^{\otimes a} \otimes \tilde \eA_s \ket{\psi_s})/\normtwo{\tilde \eA_s \ket{\psi_s}}$ with probability $\normtwo{\tilde \eA_s \ket{\psi_s}}^2$.
In this case, the ancilla register is measured in the zero state and the signal register is in the target state $\tilde \eA_s \ket{\psi_s}$, see \Cref{fig:BE}.
An inadmissible state orthogonal to the desired outcome is obtained with probability $1 - \normtwo{\tilde \eA_s \ket{\psi_s}}^2$.

Using amplitude amplification, the process must be repeated $1/\normtwo{\tilde \eA_s \ket{\psi_s}}$ times for success on average.
This makes our proposed synthesis technique probabilistic.

\section{Combining block-encodings}

We introduce two operations on block-encodings that in combination allow us to build encodings of larger operators from encodings of small operators.
The first operation creates a block-encoding of a Kronecker product of two matrices from the block-encodings of the individual matrices.
We denote a SWAP-gate on the $i$th and $j$th qubits as $\swap_j^i$.
\begin{lemma}
\label{lemma:KP}
Let $\eU_{n}$ and $\eU_{m}$ be $(\alpha,a,\epsilon_1)$- and $(\beta,b,\epsilon_2)$-block-encodings of $\eA_{s}$ and $\eA_{t}$, respectively, and define
$\eS_{n + m} = \prod_{i=1}^{s} \, \swap^{a+i}_{a+b+i}$.
Then,
\begin{myequation}\label{eq:TP}
\eS_{n + m}
\left( \eU_{n} \otimes \eU_{m} \right)
\eS_{n + m}^{\dagger}
\end{myequation}
is an $(\alpha\beta, a + b, \alpha \epsilon_2 + \beta \epsilon_1 + \epsilon_1 \epsilon_2)$-block-encoding of $\eA_{s} \otimes \eA_{t}$.
\end{lemma}

The proof of \Cref{lemma:KP} is given in \Cref{app:proof-lemma1}.
This lemma shows how two individual block-encodings can be combined to encode the Kronecker product of two matrices.
The method requires no additional ancilla qubits and the approximation error scales as a weighted sum of the individual errors up to first order.
The operation requires only $2s$ additional SWAP operations.

\begin{figure}[hbtp]
\centering
\subfloat{%
\ifnum\dowordcount=0
\parbox{\columnwidth}{%
\figname{circuit-kronprod}%
\begin{tikzpicture}[on grid]
\clip (-13mm,3mm) rectangle (60mm,-32mm);
\pgfkeys{/myqcircuit, layer width=9mm, row sep=3mm, source node=qwsource, gate offset=0}
\newcommand{\qwstart}{1}
\newcommand{\qwend}{5}
\qwire[start node=qwsource, end node=qw1e, style=thin,            index=1, start layer=\qwstart, end layer=\qwend ]
\qwire[start node=qw2s,       end node=qw2e, style=thin,            index=2, start layer=\qwstart, end layer=\qwend]
\qwire[start node=qw3s,       end node=qw3e, style=thin,            index=3, start layer=\qwstart, end layer=\qwend]
\qwire[start node=qw4s,       end node=qw4e, style=markedwire, index=4, start layer=\qwstart, end layer=\qwend]
\qwire[start node=qw5s,       end node=qw5e, style=markedwire, index=5, start layer=\qwstart, end layer=\qwend]
\qwire[start node=qw6s,       end node=qw6e, style=markedwire, index=6, start layer=\qwstart, end layer=\qwend]

\qwire[start node=qw8s,       end node=qw8e, style=thin, index=8, start layer=\qwstart, end layer=\qwend]
\qwire[start node=qw9s,       end node=qw9e, style=thin, index=9, start layer=\qwstart, end layer=\qwend]
\qwire[start node=qw10s,       end node=qw10e, style=markedwire, index=10, start layer=\qwstart, end layer=\qwend]
\qwire[start node=qw11s,       end node=qw11e, style=markedwire, index=11, start layer=\qwstart, end layer=\qwend]

\swapgate[layer=1, first index=4, second index=6]
\swapgate[layer=1.5, first index=5, second index=8]
\swapgate[layer=2, first index=6, second index=9]
\multiqubit[label=$\eU_{n}$, layer=3, start index=1, stop index=6, node=Un1]
\multiqubit[label=$\eU_{m}$, layer=3, start index=8, stop index=11, node=Un2]
\swapgate[layer=4, first index=6, second index=9]
\swapgate[layer=4.5, first index=5, second index=8]
\swapgate[layer=5, first index=4, second index=6]

\draw[decorate,decoration={brace,mirror,amplitude=5pt},thick]
        ($(qwsource)$)
        to node[midway,left,xshift=-2mm] (bracket) {$a$}
        ($(qw3s)$);
\draw[decorate,decoration={brace,mirror,amplitude=5pt},thick]
        ($(qw4s)$)
        to node[midway,left,xshift=-2mm] (bracket) {$s$}
        ($(qw6s)$);
\draw[decorate,decoration={brace,mirror,amplitude=5pt},thick]
        ($(qw8s)$)
        to node[midway,left,xshift=-2mm] (bracket) {$b$}
        ($(qw9s)$);
\draw[decorate,decoration={brace,mirror,amplitude=5pt},thick]
        ($(qw10s)$)
        to node[midway,left,xshift=-2mm] (bracket) {$t$}
        ($(qw11s)$);
\node [node distance=11mm, left of=\sourcenode] {(a)};
\end{tikzpicture}}
\else
  \parbox{\columnwidth}{~}
\fi
}

\subfloat{%
\ifnum\dowordcount=0
\figname{circuit-Up}%
\begin{tikzpicture}[on grid]
\clip (-28.5mm,3mm) rectangle (44.5mm,-9mm);
\pgfkeys{/myqcircuit, layer width=11mm, row sep=3mm, source node=qwsource, gate offset=0}
\newcommand{\qwstart}{1}
\newcommand{\qwend}{1}
\qwire[start node=qwsource,       end node=qw1e, style=thin,            index=1, start layer=\qwstart, end layer=\qwend,label=$a + b$]
\qwire[start node=qw3s,       end node=qw3e, style=markedwire, index=3, start layer=\qwstart, end layer=\qwend, label=$s + t$]

\multiqubit[label=$\eU_{p}$, layer=1.1, start index=1, stop index=3, node=Un]
\singlequbit[style=bundle, index=1, layer=0.5]
\singlequbit[style=bundle, index=3, layer=0.5]
\node[node distance=3.2*\layerwidth, right of=\sourcenode] {};
\node[node distance=23mm, left of=\sourcenode] {};
\node [node distance=26.5mm, left of=\sourcenode] {(b)};
\end{tikzpicture}
\else
  \parbox{\columnwidth}{~}
\fi
}
\caption{Block-encoding of the Kronecker product of 2 block-encoded matrices: (a) quantum circuit for $a = 3$, $s = 3$, $b = 2$, $t = 2$, and (b) equivalent multi-qubit gate $\eU_p$ with $p = n + m$.\label{fig:TPBE}}
\end{figure}
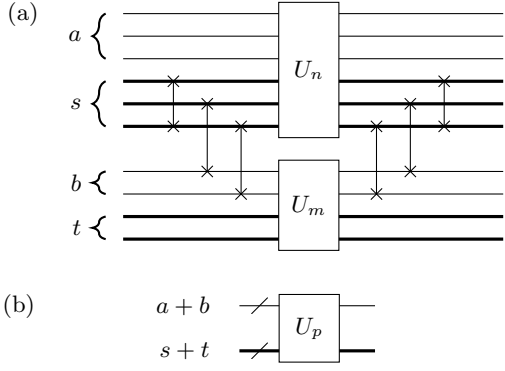

\Cref{fig:TPBE} shows the quantum circuit for a Kronecker product of block-encodings.
This reveals the observation that in order to combine block-encodings into Kronecker products, the signal qubits of the leading block-encoding have to be swapped with the ancilla qubits of the second block-encoding in such a way that the $s+t$ signal qubits become the least-significant qubits in the combined circuit and that the mutual ordering of the signal qubits is preserved.

\Cref{lemma:KP} trivially extends to Kronecker products of more than two block-encodings.
Let $\eU_{n_i}$ be $(\alpha_i,a_i,\epsilon_i)$-block-encodings of $\eA_{s_i}$
for $i \in \lbrace 1,\dots,d \rbrace$.
Define $n = \sum_i n_i$, and $S_n$ as a SWAP register that
swaps all signal qubits of each block-encoding $\eU_{n_i}$
to the least significant qubits of the $n$-qubit unitary while
preserving the mutual ordering between the signal qubits.
Then, ignoring the second order error terms,
\begin{myequation}
\label{eq:TPnd}
\eS_{n} \left( \eU_{n_1} \otimes \eU_{n_2} \otimes \cdots \otimes \eU_{n_{d}} \right) \eS_{n}^{\dagger}
\end{myequation}
is an $(\prod_i \alpha_{i}, \sum_i a_{i}, \sum_i \epsilon_{i} \prod_{k \neq i} \alpha_{k})$-block-encoding of $\eA_{s_1} \otimes \eA_{s_2} \otimes \cdots \otimes \eA_{s_d}$.
In order for the subnormalization factor and approximation error on the Kronecker product not to grow too large, the subnormalization factors of the individual block-encodings should be small enough.

The second operation used in the proposed technique constructs a block-encoding of a linear combination of block-encodings.
To this end, we review the notion of a \emph{state preparation pair of unitaries}~\cite{gisu2019}.

\begin{definition}
\label{def:SP}
Let $\exy \inC[m]$, with $\normone[]{\exy} \leq \beta$, and define $\underline{\exy} = \left[ \exy^T \, 0 \right]^T \inC[2^b]$, where $2^b \geq m$.
Then the pair of unitaries $(\eP_b,\eQ_b)$ is called a $(\beta, b,\epsilon)$-state-preparation-pair for $\exy$ if
$\eP_b \ket{0}^{\otimes b} = \ket{p}$ and $\eQ_b \ket{0}^{\otimes b} = \ket{q}$, such that
\begin{myequation}
\sum_{j=0}^{2^b-1} | \beta (p_{j}^* q_j) - \underline{y_j} | \leq \epsilon.
\end{myequation}
\end{definition}

The following lemma is a known result \cite{chwi2012}, but we provide a sharper upper bound on the approximation error compared to \cite{gisu2019}.

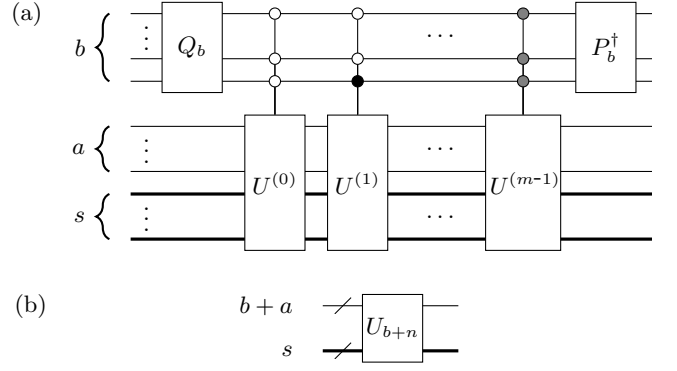
\begin{figure}[hbtp]
\centering

\subfloat{%
\ifnum\dowordcount=0
\figname{circuit-lincomb}%
\begin{tikzpicture}[on grid]
\clip (-13mm,3mm) rectangle (72mm,-32mm);      
\pgfkeys{/myqcircuit, layer width=11mm, row sep=3mm, source node=qwsource, gate offset=0}
\newcommand{\qwstart}{1}
\newcommand{\qwend}{6}
\qwire[start node=qwsource, end node=qw1e, style=thin,            index=1, start layer=\qwstart, end layer=\qwend ]
\drawdots[layer=0, index=2]
\qwire[start node=qw4s,       end node=qw4e, style=thin,           index=3, start layer=\qwstart, end layer=\qwend]
\qwire[start node=qw5s,       end node=qw5e, style=thin,           index=4, start layer=\qwstart, end layer=\qwend]
\qwire[start node=qw7s,       end node=qw7e, style=thin,           index=6, start layer=\qwstart, end layer=\qwend]
\drawdots[layer=0, index=7]
\qwire[start node=qw10s,       end node=qw10e, style=thin,       index=8, start layer=\qwstart, end layer=\qwend]
\qwire[start node=qw11s,       end node=qw11e, style=markedwire, index=9, start layer=\qwstart, end layer=\qwend]
\drawdots[layer=0, index=10]
\qwire[start node=qw14s,       end node=qw10e, style=markedwire, index=11, start layer=\qwstart, end layer=\qwend]

\multiqubit[label=$\eQ_{b}$, layer=1, start index=1, stop index=4, node=Qb]
\multiqubit[label=$\eU\p{0}$, layer=2, start index=6, stop index=11, node=U0]
\control[layer=2, index=1, style=controloff, target node=U0]
\control[layer=2, index=3, style=controloff, target node=U0]
\control[layer=2, index=4, style=controloff, target node=U0]

\multiqubit[label=$\eU\p{1}$, layer=3, start index=6, stop index=11, node=U1]
\control[layer=3, index=1, style=controloff, target node=U1]
\control[layer=3, index=3, style=controloff, target node=U1]
\control[layer=3, index=4, style=controlon, target node=U1]

\node[below right = 1*\rowsep and 4*\layerwidth of qwsource]{$\ldots$};
\node[below right = 6*\rowsep and 4*\layerwidth of qwsource]{$\ldots$};
\node[below right = 9*\rowsep and 4*\layerwidth of qwsource]{$\ldots$};

\multiqubit[label=$\,\eU\p{m\mbox{-}1}$, layer=5, start index=6, stop index=11, node=Um]
\control[layer=5, index=1, style=controlhalf, target node=Um]
\control[layer=5, index=3, style=controlhalf, target node=Um]
\control[layer=5, index=4, style=controlhalf, target node=Um]
\multiqubit[label=$\eP_{b}^{\dagger}$, layer=6, start index=1, stop index=4, node=Pb]

\draw[decorate,decoration={brace,mirror,amplitude=5pt},thick]
        ($(qwsource)$)
        to node[midway,left,xshift=-2mm] (bracket) {$b$}
        ($(qw5s)$);
\draw[decorate,decoration={brace,mirror,amplitude=5pt},thick]
        ($(qw7s)$)
        to node[midway,left,xshift=-2mm] (bracket) {$a$}
        ($(qw10s)$);
\draw[decorate,decoration={brace,mirror,amplitude=5pt},thick]
        ($(qw11s)$)
        to node[midway,left,xshift=-2mm] (bracket) {$s$}
        ($(qw14s)$);
\node [node distance=11mm, left of=\sourcenode] {(a)};  
\end{tikzpicture}
\else
  \parbox{\columnwidth}{~}
\fi
}

\subfloat{%
\ifnum\dowordcount=0
\parbox{\columnwidth}{%
\figname{circuit-Ubpn}%
\begin{tikzpicture}[on grid]
\clip (-38mm,3mm) rectangle (47mm,-8mm);      
\pgfkeys{/myqcircuit, layer width=11mm, row sep=3mm, source node=qwsource, gate offset=0}
\newcommand{\qwstart}{1}
\newcommand{\qwend}{1}
\qwire[start node=qwsource,       end node=qw1e, style=thin,            index=1, start layer=\qwstart, end layer=\qwend,label=$b + a$]
\qwire[start node=qw3s,       end node=qw3e, style=markedwire, index=3, start layer=\qwstart, end layer=\qwend, label=$s$]
\multiqubit[label=$\eU_{b+n}$, layer=1.1, start index=1, stop index=3, node=Un]
\singlequbit[style=bundle, index=1, layer=0.5]
\singlequbit[style=bundle, index=3, layer=0.5]
\node [node distance=36mm, left of=\sourcenode] {(b)};
\end{tikzpicture}}
\else
  \parbox{\columnwidth}{~}
\fi
}
\caption{Block-encoding of linear combinations of block-encodings: (a) quantum circuit where the
white control nodes are controlled on the $\ket{0}$ state, the black control nodes on the $\ket{1}$
state, and the
gray control nodes for $\eU\p{m-1}$
are controlled on either the $\ket{0}$ or $\ket{1}$ state in order to
encode the bitstring for $m-1$, and (b) equivalent multi-qubit gate.\label{fig:LCUBE}}
\end{figure}

\begin{lemma}
\label{lemma:LCU}
Let $\eB_s = \sum_{j=0}^{m-1} y_j \eA_{s}\p{j}$ be an $s$-qubit operator
and assume that $(\eP_b,\eQ_b)$ is a $(\beta,b,\epsilon_1)$-state-preparation-pair for $\exy$.
Further, let $\eU_{n}\p{j}$ be $(\alpha, a, \epsilon_2)$-block-encodings for $\eA_{s}\p{j}$ for $j \in [m]$ and
define the following select oracle
\begin{myequation}
\eW_{b+n} = \sum_{j=0}^{m-1} \ket{j}\bra{j} \otimes \eU_{n}\p{j} + \sum_{j=m}^{2^b-1} \ket{j} \bra{j} \otimes \eI_n.
\label{eq:selor}
\end{myequation}
Then,
\begin{myequation}
\eU_{b+n} = (\eP_{b}^{\dagger} \otimes \eI_a \otimes \eI_s) \, \eW_{b+n} \, (\eQ_{b} \otimes \eI_a \otimes \eI_s),
\end{myequation}
is an $(\alpha\beta, a+b,\alpha \epsilon_1 + \beta \epsilon_2)$-block-encoding of $\eB_s$.
\end{lemma}
The proof is provided in \Cref{app:proof-lemma2}.
This lemma shows that, if an efficient state preparation pair exists for the coefficient vector $\exy$,
then we can efficiently implement a linear combination of block-encodings from the individual block-encodings.
\Cref{fig:LCUBE} shows the corresponding quantum circuit.
Note that this operation requires $b$ additional ancilla qubits. The approximation error again scales as a weighted sum of the (maximum) error on the block-encodings and the error on the state-preparation pair.

The combination of \Cref{lemma:LCU} and \eqref{eq:TPnd}
shows that we can directly construct
a block-encoding of an $s$-qubit operator with the CP-like form
\begin{myequation}\label{eq:CP}
\eB_s = \sum_{j=0}^{m-1} y_j \ \eA_{s_1}\p{j} \otimes \eA_{s_2}\p{j} \otimes \dots \otimes \eA_{s_{d_j}}\p{j},
\end{myequation}
if $\sum_{i=1}^{d_j} s_i = s$ for $j \in [m]$, i.e., all terms in the sum in \eqref{eq:CP} are of the same
dimension, and if we have a block-encoding  $\eU_{n_i}\p{j}$ for each $\eA_{s_i}\p{j}$
where $j \in [m]$, and $i \in \lbrace 1,\dots,d_j \rbrace$.

To quantify the subnormalization factor, the number of ancilla qubits, and the approximation error in the block-encoding for \eqref{eq:CP}, we assume that each $\eU_{n_i}\p{j}$ is  an
$(\alpha_i\p{j},a_i\p{j},\epsilon_i\p{j})$-block-encoding for $\eA_{s_i}\p{j}$.
Let
\begin{myequation}
\alpha\p{j} = \myprod_i \alpha_{i}\p{j}, \
a\p{j} = \mysum_i a_{i}\p{j}, \
\epsilon\p{j} = \mysum_i \epsilon_{i}\p{j} \myprod_{k \neq i} \alpha_{k}\p{j},
\label{eq:TPparam}
\end{myequation}
for $j \in [m]$.
Then, using \eqref{eq:TPnd}, we can combine these into
$(\alpha\p{j},a\p{j},\epsilon\p{j})$-block-encodings
for each term in \eqref{eq:CP}.
Notice that while the number of signal qubits has to be the same for each term in the linear combination,
we do not assume the same number of ancilla qubits here.
If we define $a = \max_j a\p{j}$, then each block-encoding for $\eA_s\p{j}$ can simply be extended to
$a$ ancilla qubits by adding additional ones at the top of the register. This does not change the leading block of the unitary.
The properties of a block-encoding for \eqref{eq:CP}
under these assumptions are formalized in the following theorem.

\begin{theorem}
Let $\eB_s$ be the $s$-qubit operator in \eqref{eq:CP} with
$(\alpha\p{j},a\p{j},\epsilon\p{j})$-block-encodings of
$ \eA_{s_1}\p{j} \otimes \eA_{s_2}\p{j} \otimes \dots \otimes \eA_{s_{d_j}}\p{j}$, for $j \in [m]$, 
constructed according to \eqref{eq:TPnd} with parameters given by \eqref{eq:TPparam}.
Assume that all block-encodings are extended to $a = \max_j a\p{j}$ ancilla qubits,
$\alpha = \max_j \alpha\p{j}$,
and $\epsilon_1 =\max_j \epsilon\p{j}$.
Then, by \Cref{lemma:LCU}, we can construct a unitary
$\eU_{b+n}$ that is an $(\alpha\beta, a+b,\alpha \epsilon_2 + \beta \epsilon_1)$-block-encoding of $\eB_s$.
\label{thm:CP}
\end{theorem}

\Cref{thm:CP} follows directly from the combination of \Cref{lemma:KP}
and \Cref{lemma:LCU}.
Without loss of generality, the subnormalization factors $\alpha\p{j} \leq \alpha$ can be incorporated in
the vector $\exy$ encoding the coefficients of the linear combination.

The circuit construction can be simplified for operators with CP structure instead of CP-like structure.
The combination of the SWAP registers from \eqref{eq:TPnd} with the select oracle
in \Cref{lemma:LCU} introduces generalized Fredkin gates \cite{frto1982}.
Fredkin gates are difficult to realize experimentally \cite{onok2017}
and can be avoided if every Kronecker product of the block-encodings in the linear combination uses
the same SWAP register.
In this case, the select oracle becomes
\begin{myequation}
\eW_{b+n} = \left( \eI_b \otimes S_n \right) \tilde\eW_{b+n}
            \left( \eI_b \otimes S^{\dagger}_n \right),
\label{eq:SWAPselect}
\end{myequation}
where
\begin{myequation}
\tilde\eW_{b+n} = \sum_{j=0}^{m-1} \ket{j}\bra{j} \otimes \tilde \eU_{n}\p{j} + \sum_{j=m}^{2^b-1} \ket{j} \bra{j} \otimes \eI_n,
\end{myequation}
with $\tilde \eU_{n}\p{j} = \eU_{n_1}\p{j} \otimes \dots \eU_{n_d}\p{j}$.

\section{Discussion}

Our technique combines block-encodings of small matrices to create block-encodings of larger operators that can be represented as in \eqref{eq:CP}.
This decomposition is closely related to the CP decomposition
of a tensor \cite{koba2009} and allows for more generality.
The sizes of the individual block-encoded matrices can differ in each term of the linear combination but
they must all have the same size when combined into a Kronecker product.

Optimization algorithms, such as for example alternating least squares, have been successfully used to compute approximations to CP decompositions in many applications. Even though exact CP decompositions are NP-hard to compute in general.
The optimization algorithms can be extended to accommodate for the different sizes of block-encodings in each
of the terms and
could incorporate the flexibility in size of the terms in their objective.
They can be used as such for approximate quantum circuit synthesis.
As NISQ devices suffer from noise \cite{pres2018}, the approximate nature of algorithms for CP-like decompositions can be exploited to obtain shorter circuits for less precise decompositions with fewer terms.
Under a given noise level, the error on the approximate CP-like decomposition can be balanced with the error on the individual block-encodings to find a tradeoff with short circuit depth.

One of the major challenges with using block-encodings is the introduction of an ancilla register.
This removes the constraint of strictly unitary approximations and allows for
linear combinations, but at the same time it introduces a probabilistic nature in the synthesis process and requires that the circuit is repeatedly executed until success.
This makes our strategy related to the Repeat-Until-Succes (RUS) synthesis technique for single qubit unitaries \cite{pasv2014,boro2015}.
A RUS circuit is a block-encoding of the desired operator in combination with a set of recovery operators 
to recover the input state if a failure state is measured. 
In our work we do not consider
recovery operators and assume that the computation is repeated if a failure state is measured.

Another related work is \cite{zhzh2019}, which proposes basic linear algebra subroutines for quantum computers.
Their method relies on Hamiltonian simulation of embeddings of arbitrary matrices
and also allows to approximate the action of PLTCP-like matrices using Trotter splitting for simulating
sums and Kronecker products of matrices.

\subsection{CNOT complexity}

The asymptotic gate complexity of the resulting quantum circuit synthesis technique depends on two factors:
the number of terms $m$ in the CP-like decomposition in \eqref{eq:CP} and the gate count of each individual block-encoding in the select oracle.
If we assume that $m = \bigO(\poly(s))$,
then $b = \bigO(\polylog(s))$ and quantum circuits with $\bigO(\poly(s))$ gates
for the state-preparation unitaries always exist \cite{plbr2011}.
Also the select oracle of \Cref{lemma:LCU} can in this case be implemented with
$\bigO(\poly(s))$ gates.

We call operators that can be expressed as \eqref{eq:CP} \emph{PLTCP-like matrices}
if the linear combination consists of $\bigO(\poly(s))$ terms, 
a polylogarithmic number of terms in the matrix dimension.
PLTCP-like matrices can be synthesized with 
polylogarithmic gate complexity if each term is efficiently implementable.
The precise asymptotic complexity depends on the size of every block $\eA_{s_i}\p{j}$ and the number
of gates required for their block-encoding.

The CNOT complexity for the simplest case where $B_s$ is a PLTCP matrix with $s$ terms and where every term is
a Kronecker product of $s$ $2 \times 2$ matrices is summarized in \Cref{tab:gc}.
The CNOT complexity of the select oracle is determined from the decomposition of $2$-qubit unitaries \cite{vida2004}
and the synthesis of controlled $1$-qubit unitaries \cite{babe1995}.

\begin{table*}[t]
  \centering\ifnum\dowordcount=0
  \begin{tabular}{L{0.36\linewidth}C{0.02\linewidth}C{0.2\linewidth}C{0.15\linewidth}C{0.22\linewidth}}
    \toprule
    \textbf{Circuit element}  & \textbf{\#} & \textbf{Gates} & \multicolumn{2}{c}{\textbf{Total CNOT complexity}}\\
                                         &                    &                        & Exact & Approximate\\
    \toprule
    \emph{State preparation} $(P_{\log(s)},Q_{\log(s)})$ \cite{plbr2011} &  & & $\frac{23}{24} s$ & -- \\
    \midrule
    \emph{SWAP registers} \cite{Nielsen:2011:QCQI} & $2s$ & $\swap$ gates & $6s$ & -- \\
    \midrule
    \emph{Select oracle} & $s$ & controlled $2s$-qubit & $\Theta(11s^2 \log(s)^2 )$ & $\Theta(11 s^2 \log(s) \log(1/\epsilon)) $ \\
    \hspace{1.2em}$2s$-qubit with $\log(s)$ controls & $s$ & controlled $2$-qubit & $\Theta(11s  \log(s)^2)$ & $\Theta(11 s \log(s) \log(1/\epsilon))$ \\
    \hspace{2.4em} $2$-qubit with $\log(s)$ controls \cite{vida2004,babe1995} & 11 & controlled $1$-qubit & $\Theta(11 \log(s)^2)$ & $\Theta(11 \log(s) \log(1/\epsilon))$ \\
     \hspace{3.6em} $1$-qubit with $\log(s)$ controls \cite{babe1995} & & & $\Theta(\log(s)^2)$ & $\Theta(\log(s)\log(1/\epsilon))$ \\
    \hspace{3.6em} Toffoli with $\log(s)+1$ controls \cite{babe1995} & & & $\Theta((\log(s)+1)^2)$ & $\Theta((\log(s)+1)\log(1/\epsilon))$  \\
    \bottomrule
  \end{tabular}
  \fi
  \caption{Asymptotic CNOT complexity for a quantum circuit that block-encodes a PLTCP matrix $B_s$ with $s$ terms in the linear combination and every term a Kronecker product of $s$ $2 \times 2$ matrices.
  The third column lists the CNOT complexity for an exact synthesis of a controlled single qubit gate, the fourth column for an approximate synthesis \cite{babe1995}.}
  \label{tab:gc}
\end{table*}

For PLTCP-like matrices with more complicated structures we still maintain a $\bigO(\poly(s))$ CNOT complexity
as long as the gate complexity for the synthesis of the individual block-encodings scales at most with $\bigO(\poly(s))$.
An advantage of this method is that the synthesis of the $\bigO(\poly(s))$ small block-encoding unitaries requires fewer classical resources than the synthesis of larger blocks.
The strength of the technique lies in the ability to combine small-scale block-encodings to build larger operators.

\subsection{Examples}

We stress that unitariness of $\eB_s$ is not required because of the embedding as a block-encoding
and that even if $\eB_s$ is unitary, the individual terms in \eqref{eq:CP} clearly are not unitary.
One class of PLTCP matrices 
is the Laplace-like operators \cite{krst2014}
\begin{myequation}
\sum_{j=1}^d M\p{1} \otimes \cdots \otimes M\p{j-1} \otimes L\p{j}
 \otimes M\p{j+1} \otimes \cdots \otimes M\p{d},
\end{myequation}
and they can directly be encoded from block-encodings of the individual terms.
For example in the Laplace operator itself,
all $M\p{j}$ are identities and $L\p{j} = L$ for $j \in \lbrace 1,\dots,d \rbrace$.
In this case
we only need one block-encoding of $L$, which is repeated $d$ times, to encode the full operator.
This is an improvement over the $d^2$ block-encodings that are required in general.

\begin{figure}[b!]
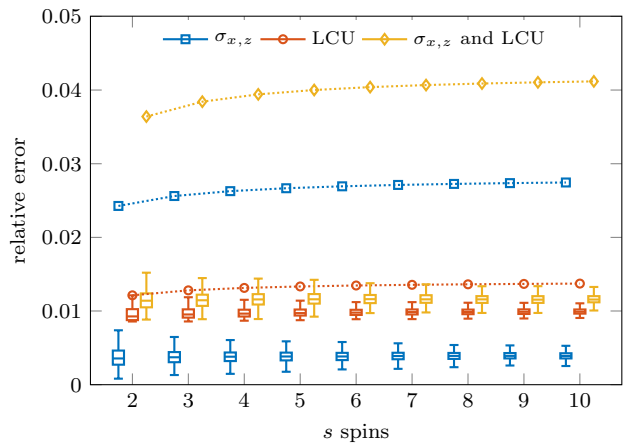

\centering
\ifnum\dowordcount=0
  \figname{block-encoding-error}%
  \myTDFIMboxplot{1}{my1DTFIM_sigma_error_1_00e-02}{my1DTFIM_LCU_error_1_00e-02}{my1DTFIM_both_error_1_00e-02}{my1DTFIM_sigma_bound_1_00e-02}{my1DTFIM_LCU_bound_1_00e-02}{my1DTFIM_both_bound_1_00e-02}
\else
  \parbox{\columnwidth}{~}
\fi
\vspace{-5pt}
\caption{Results of $1000$ simulations of $H_{\mathrm{TFIM}}$ with $2$ to $10$ spins and $h=2$.
The boxplots summarize the empirical relative errors on the block-encoding of $H_{\mathrm{TFIM}}$ under three
different error scenarios:
a $1\%$ error on the Pauli-$X$ and $Z$ gates (\textit{blue}), a $1\%$ error on the state preparation unitaries for LCU (\textit{red}), and
a $1\%$ error on both the Pauli gates and the LCU unitaries (\textit{yellow}).
The dotted lines show the theoretical upper bound on the error according to \Cref{thm:CP}. }
\label{fig:TFIM}
\end{figure}

Localized Hamiltonians are another example of PLTCP operators. 
The Hamiltonian of a transverse field Ising model (TFIM) on a one-dimensional chain of $s$ spin-$1/2$ particles is given by
\begin{myequation}
H_{\mathrm{TFIM}} = - \sum_{i=1}^{s-1} \sigma_z\p{i}\sigma_z\p{i+1}
                    - h\sum_{i=1}^{s} \sigma_x\p{i},
\label{eq:TFIM}
\end{myequation}
where $\sigma_x$ and $\sigma_z$ are the Pauli-$X$ and $Z$ matrices.
Since this Hamiltonian is a linear combination of $2s-1$ unitaries,
no ancilla qubits are required to encode the $2 \times 2$ matrices, and
no $\swap$ operations are necessary to form the Kronecker products.
The complexity of block-encoding $H_{\mathrm{TFIM}}$ lies in forming the
linear combination.
We have simulated block-encoding circuits for $H_{\mathrm{TFIM}}$
under three different error scenarios: a $1\%$ error on the $\sigma_x$ and $\sigma_z$ gates,
a $1\%$ error on the state preparation for the linear combination
of unitaries (LCU), and the combination of both.
The results are summarized in \Cref{fig:TFIM} with the theoretical
upper bound derived from \Cref{thm:CP} denoted by the dotted lines.

We observe that errors on the Pauli gates have
a smaller effect on the accuracy of the block-encoding then errors
on the state preparation unitaries.
The upper bound slightly overestimates the effect of the errors on the Pauli gates. 
This happens because the error is not uniformly
distributed over the terms in the linear combination in \eqref{eq:TFIM}.
The expected number of repetitions until success lies between $1.2$ and $1.4$ for $2$ to $10$ spins
and is not sensitive to errors.

\begin{figure}[t]
\centering\ifnum\dowordcount=0
\figname{CP-ranks}%
\begin{tikzpicture}
\begin{semilogyaxis}[%
  width=\columnwidth,%
  height=0.75\columnwidth,%
  xmin=0,xmax=20,%
  ymin=1e-8,ymax=1e0,%
  ytick={1,1e-2,1e-4,1e-6,1e-8},%
  ymajorgrids,%
  xlabel={CP rank},%
  ylabel={relative error},%
  legend style={fill=white},
]
\addplot [densely dotted, color=mycolor1, thick, mark=*, mark size=1.75pt, mark options={solid, fill=mycolor1, mycolor1}]
  table[row sep=crcr]{%
1	0.912870929175277\\
2	0.790571867523425\\
3	0.577368350920214\\
4	0.00558751336491767\\
5	0.0045384807388519\\
6	2.80346403632386e-08\\
7	1.84532706983193e-08\\
8	1.47790014360427e-08\\
};
\addlegendentry{$s=3$}

\addplot [densely dotted, color=mycolor2,thick, mark=square*, mark size=1.75pt, mark options={solid, fill=mycolor2, mycolor2}]
  table[row sep=crcr]{%
1	0.942809041582063\\
2	0.866026902713135\\
3	0.745365378113469\\
4	0.577368307847791\\
5	0.471425520620886\\
6	0.33336448692263\\
7	0.00444207126766672\\
8	0.0037159743365753\\
9	2.45678429295726e-08\\
10	2.51778848274609e-08\\
11	2.16845164690289e-08\\
};
\addlegendentry{$s=4$}

\addplot [densely dotted, color=mycolor3, thick, mark=triangle*, mark options={solid, fill=mycolor3, mycolor3}]
  table[row sep=crcr]{%
1	0.957427107756338\\
2	0.901388900005301\\
3	0.853913707438316\\
4	0.763769457617775\\
5	0.645509327896053\\
6	0.559032712075644\\
7	0.408280191818251\\
8	0.00561430217533745\\
9	0.00569672753367535\\
10	0.00519014475293054\\
11	0.00321860986695679\\
12	3.28781531000363e-08\\
13	8.36785119565928e-08\\
14	2.08111688155504e-08\\
};
\addlegendentry{$s =5$}

\addplot [densely dotted, color=mycolor4, thick, mark=diamond*, mark options={solid, fill=mycolor4, mycolor4}]
  table[row sep=crcr]{%
1	0.966091783079296\\
2	0.921955327829156\\
3	0.856353735565298\\
4	0.83666189260927\\
5	0.730304761220318\\
6	0.670830706733985\\
7	0.568781787529541\\
8	0.447284304896413\\
9	0.365190221903748\\
10	0.258302716951622\\
11	0.00578156436525717\\
12	0.00539778590169307\\
13	0.00490251968239922\\
14	0.0035782995680635\\
15	6.91993065415734e-08\\
16	1.02958716490896e-07\\
17	1.6167533500702e-07\\
};
\addlegendentry{$s=6$}

\addplot [color=black, line width=1.2pt, draw=none, mark size=4.5pt, mark=o, mark options={solid, black}, forget plot]
  table[row sep=crcr]{%
6	2.80346403632386e-08\\
};
\addplot [color=black, line width=1.2pt, draw=none, mark size=4.5pt, mark=o, mark options={solid, black}, forget plot]
  table[row sep=crcr]{%
9	2.45678429295726e-08\\
};
\addplot [color=black, line width=1.2pt, draw=none, mark size=4.5pt, mark=o, mark options={solid, black}, forget plot]
  table[row sep=crcr]{%
12	3.28781531000363e-08\\
};
\addplot [color=black, line width=1.2pt, draw=none, mark size=4.5pt, mark=o, mark options={solid, black}, forget plot]
  table[row sep=crcr]{%
15	6.91993065415734e-08\\
};
\end{semilogyaxis}
\end{tikzpicture}%
\fi
\vspace{-5pt}
\caption{Compression of the CP rank with tensor toolbox \cite{TTB_Software}
of the Heisenberg isotropic
 antiferromagnetic Hamiltonian $H_{\mathrm{XYZ}}$ for $s =3, \ldots, 6$ spins.
 The CP rank of the exact decomposition, \eqref{eq:HIAM},
 is circled.}
\label{fig:HIAM}
\end{figure}
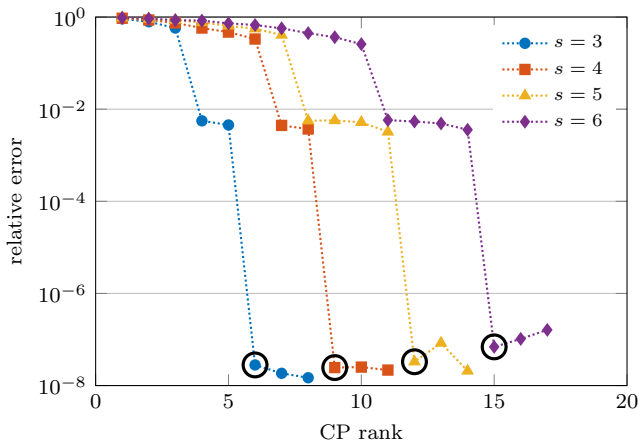

The Hamiltonian for the spin-1 Heisenberg model is equal to
\begin{myequation}
H_{\mathrm{XYZ}} = \sum_{i=1}^{s-1} X\p{i}X\p{i+1} + Y\p{i}Y\p{i+1} + Z\p{i}Z\p{i+1},
\label{eq:HIAM}
\end{myequation}
where $X, Y,$ and $Z$ are the spin-1 generators of SU(2).%
These $3 \times 3$ matrices can be embedded in $4 \times 4$ matrices by zero-padding and block-encoded
in $2$ signal qubits and $1$ ancilla qubit.
In order to compress the CP rank, 
we have \emph{tensorized} $H_{\mathrm{XYZ}}$ to an $s$-way $9 \times 9 \times \cdots \times 9$ array
and numerically computed an approximate CP decomposition using the alternating least squares
algorithm from tensor toolbox \cite{TTB_Software}.
The results for $3$ to $6$ spins are shown in \Cref{fig:HIAM}.

We observe that the relative error on the approximation of the Hamiltonian
decreases with increasing CP rank. A stagnation occurs at the exact CP rank of
the operator, signaling convergence. If an approximation with a relative error of
$1\%$ is sufficient, a CP rank reduction of $20\%-30\%$ can be achieved.
This directly translates to shorter quantum circuits as each term appears in the
select oracle. 
For example, in the case $s=4$ it also leads to a reduction
in ancilla qubits: the exact expression
is a linear combination of $9$ terms, requiring $4$ ancilla qubits for encoding the linear
combination, and this can be compressed to $7$ terms, or only $3$ ancilla qubits.

\section{Conclusions}

In this paper we showed how block-encodings of small matrices, which are easier to synthesize, can be combined together to create block-encodings of larger operators with CP-like structure.
Under the assumption of $\bigO(\poly(s))$ terms in the decomposition and small individual block-encodings, this scheme has a polynomial dependence on the number of signal qubits both for gate complexity and ancilla qubits.
We reviewed three examples of PLTCP matrices, showed that the CP rank can be compressed if a larger
approximation error is acceptable and found that the circuits behave well under errors.

Further research is required to study the class of operators with PLTCP-like structure and operators that can be well-approximated in this form.
The modification of optimization algorithms for CP decompositions \cite{koba2009} to admit decompositions like \eqref{eq:CP} is another interesting research direction.

\ifnum\dowordcount=0

\begin{acknowledgments}
This work was supported by the
Laboratory Directed Research and Development Program
of Lawrence Berkeley National Laboratory under
U.S. Department of Energy Contract No. DE-AC02-05CH11231.
\end{acknowledgments}

\fi

\bibliographystyle{apsrev4-2}
\bibliography{references}

\ifnum\dowordcount=0

\onecolumngrid
\appendix

\section{Proof of \cref{lemma:KP}}
\label{app:proof-lemma1}

\emph{Proof.}
From \Cref{def:BE} and the mixed-product property of the Kronecker product $(A \otimes B)(C \otimes D) = (AC) \otimes (BD)$, we obtain
\begin{myequation}\label{eq:permutedTP}
\tilde \eA_{s} \otimes \tilde \eA_{t} =
\left(\bra{0}^{\otimes a} \otimes \eI_{s} \otimes \bra{0}^{\otimes b} \otimes \eI_{t} \right)
\left( \eU_{n} \otimes \eU_{m} \right)
\left( \ket{0}^{\otimes a} \otimes \eI_{s} \otimes \ket{0}^{\otimes b} \otimes \eI_{t} \right).
\end{myequation}
The Kronecker product $\tilde \eA_{s} \otimes \tilde \eA_{t}$ is encoded in $\eU_{n} \otimes \eU_{m}$, but not as the leading principal block.
We use the property,
\begin{myequation*}
\swap^1_2 \left(\eI_1 \otimes \ket{0}\right)  = \ket{0} \otimes \eI_1,
\end{myequation*}
to show that $\eS_{n + m}$ recovers the correct order by swapping
the $s$ signal qubits:
\begin{myalign*}
\eS_{n + m}
\left(  \ket{0}^{\otimes a} \otimes \eI_{s} \otimes \ket{0}^{\otimes b} \otimes \eI_{t} \right)
\ & = \
 \prod_{i=1}^{s} \swap^{a+i}_{a+b+i}
\left(  \ket{0}^{\otimes a} \otimes \eI_{s} \otimes \ket{0}^{\otimes b} \otimes \eI_{t} \right),\\
\ & = \
 \prod_{i=1}^{s-1} \swap^{a+i}_{a+b+i} \, \swap^{a+s}_{a+b+s}
\left(  \ket{0}^{\otimes a} \otimes \eI_{s} \otimes \ket{0}^{\otimes b} \otimes \eI_{t} \right),\\
\ & = \
\prod_{i=1}^{s-1} \swap^{a+i}_{a+b+i}
\left(  \ket{0}^{\otimes a} \otimes \eI_{s-1} \otimes \ket{0}^{\otimes b} \otimes \eI_1 \otimes \eI_{t} \right),\\
\ & = \
\dots\\
\ & = \
\ket{0}^{\otimes a} \otimes \ket{0}^{\otimes b}\otimes \eI_{s} \otimes \eI_{t}.
\end{myalign*}
Taking the Hermitian conjugate yields
\begin{myequation*}
\left(\bra{0}^{\otimes a} \otimes \eI_{s} \otimes \bra{0}^{\otimes b} \otimes \eI_{t} \right)
\eS_{n + m}^{\dagger}
\ = \
\bra{0}^{\otimes a} \otimes \bra{0}^{\otimes b}  \otimes \eI_{s} \otimes \eI_{t}.
\end{myequation*}
Combining this with \eqref{eq:permutedTP} shows
\begin{myalign*}
\tilde \eA_{s} \otimes \tilde \eA_{t}
\ & = \
\left(\bra{0}^{\otimes a} \otimes \eI_{s} \otimes \bra{0}^{\otimes b} \otimes \eI_{t} \right)
\eS_{n+m}^{\dagger} \eS_{n+m} \left( \eU_{n} \otimes \eU_{m} \right)
\eS_{n+m}^{\dagger} \eS_{n+m}
\left( \ket{0}^{\otimes a} \otimes \eI_{s} \otimes \ket{0}^{\otimes b} \otimes \eI_{t} \right),\\
\ & = \
\left(\bra{0}^{\otimes a} \otimes \bra{0}^{\otimes b}  \otimes \eI_{s} \otimes \eI_{t} \right)
\eS_{n+m} \left( \eU_{n} \otimes \eU_{m} \right)\eS_{n+m}^{\dagger}
\left( \ket{0}^{\otimes a} \otimes \ket{0}^{\otimes b}\otimes \eI_{s} \otimes \eI_{t} \right),
\end{myalign*}
such that \eqref{eq:TP} has $\tilde \eA_{s} \otimes \tilde \eA_{t}$
as principal leading block.
The subnormalization and approximation error of $\tilde \eA_{s} \otimes \tilde \eA_{t}$ satisfy:
\begin{myalign*}
\normtwo[\big]{\eA_{s} \otimes \eA_{t}
               - \alpha \beta \tilde \eA_{s} \otimes \tilde \eA_{t}}
 &\leq
  \normtwo[\big]{ \big( \alpha \tilde \eA_{s} + \epsilon_1 \eI_{s} \big)
          \otimes \big( \beta \tilde \eA_{t} + \epsilon_2 \eI_{t} \big)
              - \alpha \tilde \eA_{s} \otimes \beta \tilde \eA_{t} }, \\
 &=
  \normtwo[\big]{ \alpha \tilde \eA_{s} \otimes \epsilon_2 \eI_{t}
                + \epsilon_1 \eI_{s} \otimes \beta \tilde \eA_{t}
                + \epsilon_1 \eI_{s} \otimes \epsilon_2 \eI_{t} }, \\
 &\leq \alpha \epsilon_2 \normtwo[\big]{\tilde\eA_{s}}
     + \beta \epsilon_2 \normtwo[\big]{\tilde\eA_{t}}
     + \epsilon_1 \epsilon_2, \\
 &\leq \alpha \epsilon_2 + \beta \epsilon_1 + \epsilon_1 \epsilon_2,
\end{myalign*}
where we used that $\normtwo{\eA_{s}} \leq \alpha \normtwo{\tilde \eA_{s}} + \epsilon_1$, and $\normtwo{\tilde \eA_{s}} \leq 1$ and analogous results for $\tilde \eA_t$.
This completes the proof.
\hfill$\square$

\section{Proof of \cref{lemma:LCU}}
\label{app:proof-lemma2}

\emph{Proof.}
We have that the leading $s$-qubit block of $\eU_{b+n}$ is given by
\begin{myalign*}
\tilde \eB_s = \ &
\big( \bra{0}^{\otimes b}  \otimes \bra{0}^{\otimes a} \otimes \eI_s \big)
\ \eU_{b+n} \
\big( \ket{0}^{\otimes b} \otimes \ket{0}^{\otimes a} \otimes \eI_s \big),\\
=\ &
\big( \bra{0}^{\otimes b}  \otimes \bra{0}^{\otimes a} \otimes \eI_s \big)
\ \big(\eP_{b}^{\dagger} \otimes \eI_a \otimes \eI_s) \ \eW_{b+n} \ (\eQ_{b} \otimes \eI_a \otimes \eI_s \big) \
\big( \ket{0}^{\otimes b} \otimes \ket{0}^{\otimes a} \otimes \eI_s \big),\\
=\  &
\big(\bra{0}^{\otimes b}  \eP_{b}^{\dagger} \otimes \bra{0}^{\otimes a}  \otimes \eI_s \big)
\ \eW_{b+n} \
\big(\eQ_{b}\ket{0}^{\otimes b}  \otimes \ket{0}^{\otimes a}  \otimes \eI_s\big),\\
=\  &
\big(\bra{p} \otimes \bra{0}^{\otimes a} \otimes \eI_s \big)
\ \eW_{b+n} \
\big(\ket{q} \otimes \ket{0}^{\otimes a}  \otimes \eI_s\big).
\end{myalign*}
Plugging in the expression for the select oracle, \eqref{eq:selor}, this yields
\begin{myalign*}
\tilde \eB_s & =
\sum_{j=0}^{m-1}  \braket {p | j} \braket{j | q} \otimes  (\bra{0}^{\otimes a} \otimes \eI_s) \eU_{n}\p{j} (\ket{0}^{\otimes a}  \otimes \eI_s)
\ + \
\sum_{j=m}^{2^b-1}  \braket {p | j} \braket{j | q} \otimes \bra{0}^{\otimes a} \ket{0}^{\otimes a} \otimes \eI_s,\\
& =
\sum_{j=0}^{m-1}  p_{j}^* q_j \, \tilde \eA_{s}\p{j}
\ + \
\sum_{j=m}^{2^b-1}  p_{j}^* q_j  \, \eI_s.
\end{myalign*}
By \Cref{def:BE} and \Cref{def:SP}, we get that
\begin{myalign*}
\normtwo[\Big]{\eB_s - \alpha \beta \tilde \eB_s}
& =
\normtwo[\Bigg]{
\sum_{j=0}^{m-1} y_j \eA_{s}\p{j} -
\alpha \beta \sum_{j=0}^{m-1}   p_{j}^* q_j \, \tilde \eA_{s}\p{j}
\ - \  \alpha \beta
\sum_{j=m}^{2^b-1}   p_{j}^* q_j \, \eI_s}, \\
& =
\normtwo[\Bigg]{
\sum_{j=0}^{m-1} y_j \eA_{s}\p{j} - \alpha \beta  p_{j}^* q_j \, \tilde \eA_{s}\p{j}
\ - \
\alpha \sum_{j=m}^{2^b-1} \beta  p_{j}^* q_j \, \eI_s},\\
& \leq
\alpha \epsilon_1 +
\normtwo[\Bigg]{\sum_{j=0}^{m-1} \underline{y_j} (\eA_{s}\p{j} - \alpha  \tilde \eA_{s}\p{j})} +
\alpha \normtwo[\Bigg]{\sum_{j=m}^{2^b-1} \underline{y_j}  \, \eI_s}, \\
& \leq
\alpha \epsilon_1 + \beta \epsilon_2.
\end{myalign*}
The penultimate inequality approximates all $\beta p_{j}^*q_j$ terms by $\underline{y_j}$ in the two sums. The error of each individual approximation is bounded by $\epsilon_1$, such that the total error is bounded from above by $\alpha \epsilon_1$ as $\normtwo{\tilde \eA_{s}\p{j}} \leq 1$ and $\normtwo{\eI_s} = 1$.
The last term in the penultimate line is equal to zero by \Cref{def:SP}.
The final equality directly follows from the block-encoding property and $\normone[]{\exy} \leq \beta$.
\hfill$\square$

\fi

\end{document}